\documentclass[conference, twocolumn]{IEEEtran}
\usepackage{color}
\usepackage{graphicx}
\usepackage{amsmath}
\usepackage{amsthm}
\usepackage{cite}
\usepackage{commath}
\usepackage[a4paper,margin=1.5cm]{geometry}
\usepackage{epsfig,epstopdf,amssymb,enumerate} 
\usepackage{amssymb,amsmath,multirow, array}
\usepackage[caption=false,font=footnotesize]{subfig}

\newcommand{\um}{\mathrm{m}}

\begin{document}

\title{A Study of Delay Drifts on Massive MIMO Wideband Channel Models}
\author{Carlos F. L\'{o}pez and Cheng-Xiang Wang\\

\small{Institute of Sensors, Signals and Systems, Heriot-Watt University, Edinburgh EH14 4AS, UK.} \\
Email: \{c.f.lopez, cheng-xiang.wang\}@hw.ac.uk }
\maketitle

\begin{abstract} In this paper, we study the effects of the variations of the propagation delay over large-scale antenna-arrays used in massive multiple-input multiple-output (MIMO) wideband communication systems on the statistical properties of the channel. Due to its simplicity and popularity, the Elliptical geometry-based stochastic channel model (GBSM) is employed to demonstrate new non-stationary properties of the channel in the frequency and spatial domains caused by the drift of delays. In addition, we show that the time of travel of multi-path components (MPCs) over large-scale arrays may result in overlooked frequency and spatial decorrelation effects. These are theoretically demonstrated by deriving the space-time-frequency correlation functions (STFCFs) of both narrowband and wideband Elliptical models. Closed-form expressions of the array-variant frequency correlation function (FCF), power delay profile (PDP), mean delay, and delay spread of single- and multi-confocal Elliptical models are derived when the angles of arrival (AOAs) are von Mises distributed. In such conditions, we find that the large dimensions of the antenna array may limit the narrowband characteristic of the single-ellipse model and alter the wideband characteristics (PDP and FCF) of the multi-confocal Elliptical channel model. Although we present and analyze numerical and simulation results for a particular GBSM, similar conclusions can be extended to other GBSMs. 

\end{abstract}
\section{Introduction}
MIMO technologies using large-scale antenna-arrays, i.e., massive MIMO, are considered promising technologies to cope with the increasing demand of data rate and reliable communications in the future. Arrays counting on hundreds or even thousands of antenna-elements can provide large spectral efficiency and reliability through spatial multiplexing, spatial modulation, diversity, and other MIMO techniques \cite{Persson2012,Larsson2013,Wang14ComMag, Björnson2016}. 

However, to achieve some of the benefits of massive MIMO such us increased angular resolution and diversity, antenna-elements of the array cannot be packed as much as desired \cite{Tse2005}. Thus, massive MIMO deployments often result in antenna arrays spanning long distances beyond the stationary region of the channel. Measurements have demonstrated that this may lead to the so called near-field effects and non-stationary properties of the channel along the array \cite{Gao2012,Gao2013,Li2015,Payami2012,Gao2015,Gao2015a}. These effects include array-variant angles of arrival (AOAs), propagation delays, received power, MPCs (dis)apperance and others. 

In order to efficiently assess and design new communication systems that rely on realistic properties of the channel, existing models designed for former communication systems \cite{SCM, WINNERII-models,Verdone2011,IMT-A} have been improved or redesigned accounting for these new non-stationary effects over the array \cite{Gao2013,Wu2014, Lopez16, Lopez2018, Wu2015a,Quadriga2014,Metis15,Wang2016b}. State-of-the-art massive MIMO channel models considered near-field effects by including high-order wavefronts such as spherical or parabolic wavefronts. Besides, they include clusters visibility over the array through visibility regions \cite{Gao2013}, (dis)appearance or (re)appearance Markov processes \cite{Wu2014,Wu2015a,Lopez16,Lopez2018}. Smooth variations over the array of the clusters' average power were introduced in \cite{Lopez16} and \cite{Lopez2018} by employing lognormal spatial processes. However, previous works neglected \cite{Wu2014, Lopez16,Wu2015a} or did not studied \cite{Quadriga2014,Metis15} the effects of the time-delay required by signals traveling over a large array on the channel's statistical properties. Although this \emph{delay drift} over the array is present in conventional small arrays, it is usually neglected due to the small bandwidth and array dimensions used in former MIMO communication systems. 

In this paper, we will prove that delay drifts render the channel's transfer function non-stationary over the array and introduce overlooked decorrelation effects in the frequency domain as the array grows large. 
As long as these decorrelation effects may be regarded as artifacts of the channel model, we will show that the drift of delays may lead existing GBSMs towards their limits of operation. In addition, we will also demonstrate that delay drifts may result into spatial non-stationary properties of the channel over the frequency domain when the bandwidth considered is very large. 

Since the drift of delays is present regardless of the wavefront considered, e.g., plane or spherical, and we have not included cluster disappearance or shadowing over the array in this study, the effects mentioned are not caused by the type of wavefront, nor the cluster evolution along the array. Hence, to the best of the authors' knowledge, this can be considered a novel effect that has not been studied so far in literature separately. Moreover, as these effects depend on the geometrical configuration of the scatterers in the channel, e.g., Elliptical or One-Ring, and they may have an impact on the performance of new massive MIMO communication systems, hence they need to be considered for properly extending existing MIMO GBSMs to large-scale antenna-arrays. 

This paper is organized as follows: in Section~\ref{sec:model}, the elliptical GBSMs is introduced, including the delay drift over the array in the channel impulse response (CIR). In Section~\ref{sec:stat_prop}, we derive the array-variant STFCF, PDP, mean delay, and delay spread of the channel and study the effects of the delay drift on them. In Section~\ref{sec:results}, we present and analyze numerical and simulation results of the statistical properties derived in Section~\ref{sec:stat_prop}. Finally, conclusions are drawn in Section~\ref{sec:conclusions}.
\section{Delay Drift in the Elliptical GBSM}
\label{sec:model}
\begin{figure}[t]
\centering
 \includegraphics[width=0.48\textwidth]{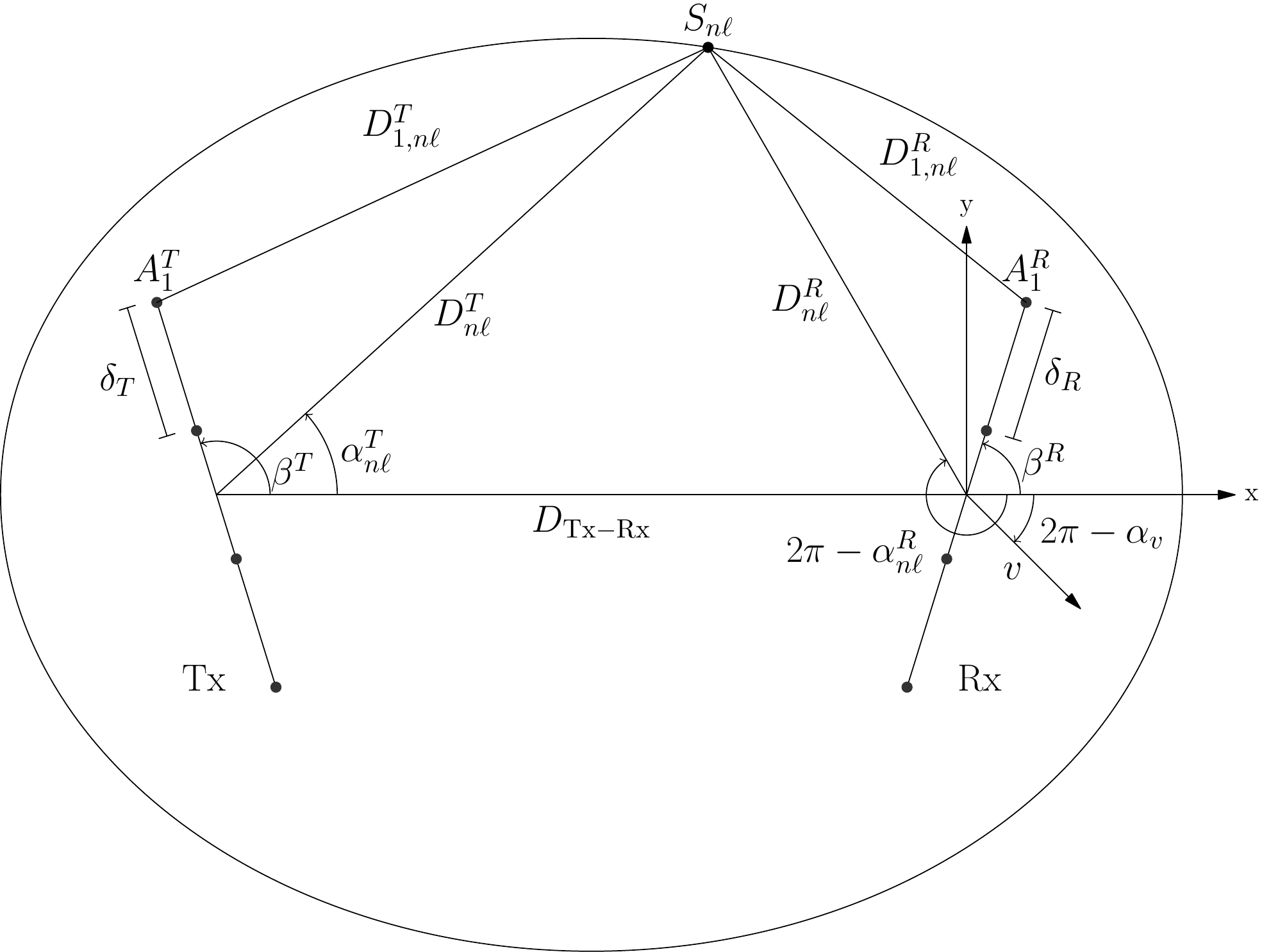} 
  \caption{Elliptical scattering model for an $M_T \times M_R$ MIMO channel.}
\label{fig:channel_model}
\end{figure}

Due to the simplicity and popularity of the model, we employ the geometrical Elliptical scattering model in this study. This model is widely used for theoretical and practical purposes (see \cite{Patzold2012}). For the reader's convenience, the model is depicted in Fig.~\ref{fig:channel_model}, where only one ellipse is depicted for clarity.
The transmitter (Tx) and receiver (Rx) are equipped with uniform linear arrays (ULAs) tilted $\beta_T$ and $\beta_R$ with respect to (wrt) the x-axis, respectively. The transmitting (receiving) ULA is formed by $M_T$ ($M_R$) antenna elements with inter-element spacing $\delta_T$ ($\delta_R$). The $p$th transmitting and $q$th receiving antenna-elements are denoted as $A^T_p$ and $A^R_q$, respectively.
As in most massive MIMO systems, the size of the arrays is large compared to the wavelength, i.e, $(M_T-1)\delta_T  \gg \lambda_0$ and $(M_R-1)\delta_R\gg \lambda_0$ with $\lambda_0$ the carrier wavelength. Moreover, the motion of the Rx is uniform with constant speed $v$ forming an angle $\alpha_v$ wrt the x-axis. 

The center of the arrays are located at the foci of a set of $\mathcal{L}$ concentric ellipses all with the same focal distance $f'$, and semi-mayor and semi-minor axes lengths $a'_\ell$ and $b'_\ell$, respectively, with $\ell=1,2,\dots,\mathcal{L}$. 
Every ellipse counts on $N_\ell$ scatterers randomly distributed over its perimeter, denoted as $S_{n\ell}$ in the figure. Scatterers of the $\ell$th ellipse are defined by their angle of departure (AOD) $\alpha_{n\ell}^T$ and angle of arrival (AOA) $\alpha_{n\ell}^R$ measured from the Tx and Rx array centers, respectively. In this model, it is usually considered that the signal radiated by $A^T_p$ is bounced only once by $S_{n\ell}$ before it reaches $A^R_q$. Thus, this signal experiences a total propagation delay $\tau_{qp,n\ell}=(D_{p,n\ell}^T+D_{q,n\ell}^R)/c_0$, where $c_0$ denotes the speed of light, and $D_{p,n\ell}^T$ and $D_{q,n\ell}^R$ denote the distances from $A^T_p$ to $S_{n\ell}$ and that from $S_{n\ell}$ to $A^R_q$, respectively.

\subsection{Channel Impulse Response (CIR) of the Channel Model}
The wideband CIR between the $A^T_p$ and $A^R_q$ is represented by the channel matrix ${\ {\bf H}}(t,\tau)=[h_{qp}(t,\tau)]_{M_R \times M_T}$ with $p=1,2, \dots, M_T$ and $q=1,2 \dots, M_R$. The CIR is calculated as
\begin{equation}
\begin{split}
h_{qp}(t,\tau) =&  \sum_{\ell=1}^{\mathcal{L}} c_\ell \lim_{N_\ell \to \infty} \frac{1}{\sqrt{N_\ell}} \\
&\times \sum_{n=1}^{N_\ell} a_{p,n\ell}  b_{q,n\ell} \: e^{j ( 2\pi f_{n\ell} t + \theta_{n\ell} ) } \delta(\tau - \tau_{qp,n\ell})
\label{eq:CIR}
\end{split}
\end{equation}
with $j=\sqrt{-1}$, $c_\ell$ denoting the gain of the $\ell$th path and $\theta_{n\ell}$ denoting the phase shift produced by the $n$th scatterer in the $ \ell$th ellipse, which is usually considered uniformly distributed over the interval $(0, 2\pi]$, i.e., $\theta_{n\ell} \sim \mathcal{U}(0,2\pi)$. Here, we assume that the total received power is normalized such that $\sum_{\ell=1}^{\mathcal{L}} c_\ell^2 =1$. The Doppler frequencies $f_{n\ell}$ and sub-path gains $a_{p,n\ell}, b_{q,n\ell}$ can be calculated as
\begin{eqnarray}
f_{n\ell} &=& f_{\text{max}} \cos ( \alpha_{n\ell}^R -\alpha_v) \\
a_{p,n\ell}  &=& e^{j k \delta_p \cos( \alpha_{n\ell}^T -\beta_T) }\\
b_{q,n\ell}  &=& e^{j k\delta_q \cos( \alpha_{n\ell}^R -\beta_R) }
\end{eqnarray}
where $f_\text{max}=v/\lambda_0$ and $k=2\pi/\lambda_0$ denote the maximum Doppler frequency and the wavenumber, respectively. The  terms $\delta_p = (M_T-2p+1)\delta_T/2$ and $\delta_q = (M_R-2q+1)\delta_R/2$ denote the distances from $A^T_p$ and $A^R_q$ to the center of the Tx and Rx arrays, respectively. Note that the propagation delay $\tau_{qp,n\ell}$ from $A^T_p$ to $A^R_q$ via the $n$th scatterer of the $\ell$th path in \eqref{eq:CIR} depends on the antennas $q$ and $p$, and the scatterer. In order to simplify the analysis of the delay drift over the array, we will consider a first-order approximation as
\begin{equation}
\tau_{qp,n\ell} = \tau_{0,\ell} - \tau_p \cos( \alpha_{n\ell}^T -\beta_T) - \tau_q \cos( \alpha_{n\ell}^R -\beta_R)
\label{eq:delay}
\end{equation}
where $\tau_{0,\ell}$ is the reference delay of the $\ell$th path from the transmitting to the receiving array centers, $\tau_i = \delta_i/c_0$, for $i=\{p,q\}$, denotes the propagation delay from the center of the corresponding array to the $i$th antenna element. Therefore, the terms $\tau_p \cos( \alpha_{n\ell}^T -\beta_T)$ and $\tau_q \cos( \alpha_{n\ell}^R -\beta_R)$ in \eqref{eq:delay} are employed to model the delay difference experienced by the signal radiated from $A^T_p$ and received by $A^R_q$ wrt $\tau_{0,\ell}$. Note that, as $\delta_p/c_0$ and $\delta_q/c_0$ are small in conventional systems, previous models assumed a constant delay that is independent of the antenna-element as $\tau_{qp,n\ell} = \tau_{0,\ell}$. 

The geometrical relationship between the AOA and the AOD in the elliptical model is \cite{Patzold2012}
\begin{equation}
\alpha _{n\ell}^T=\left\{\begin{matrix}
g(\alpha _{n\ell}^R ) & \mathrm{if} &0 < \alpha _{n\ell}^R  \leqslant \alpha _{0,\ell} \\ 
g(\alpha _{n\ell}^R ) + \pi & \mathrm{if} & \alpha _{0,\ell}  < \alpha _{n\ell}^R  \leqslant 2\pi  - \alpha _{0,\ell} \\ 
g(\alpha _{n\ell}^R ) + 2\pi & \mathrm{if} & 2\pi  - \alpha _{0,\ell}  < \alpha _{n\ell}^R  \leqslant 2\pi
\end{matrix}\right.
\label{eq:alpha_r_t}
\end{equation}
where
\begin{equation}
g(\alpha _{n\ell}^R ) = \arctan \left[ {{ \frac{(k_\ell^2  - 1)\sin (\alpha _{n\ell}^R )}{2k_\ell  + (k_\ell^2  + 1)\cos (\alpha _{n\ell}^R )}}} \right]
 \label{f_AoA_AoD}
\end{equation}
and
\begin{equation}
\alpha_{0,\ell}  = \pi  - \arctan \left( \frac{k_\ell^2  - 1}{2k_\ell }\right)
\end{equation}
where the term $k_\ell  = a_\ell/f$ denotes the inverse of the eccentricity of the $l$-th ellipse.

In the following, to describe both isotropic and non-isotropic scattering, the AOA is modeled by the flexible von Mises distribution. The probability density function of this distribution is defined as
\begin{equation}
f(x) = \frac{1}{2\pi I_0(\kappa)} e^{\kappa \cos( x - \mu )}
\end{equation}
for $x \in (0,2\pi]$. The term $I_0(\cdot)$ denotes the zero-order modified Bessel function of the first kind, $\mu \in (0, 2\pi]$ denotes the mean angle and $\kappa \ge 0$ the concentration parameter that controls the angular spread around $\mu$. The von Mises distribution reduces to $\mathcal{U}(0,2\pi)$ for $\kappa = 0$ and it approximates the Gaussian distribution with standard deviation $\sigma=2/\sqrt{\kappa}$ for large values of $\kappa$ \cite{Patzold2012}.
\subsection{Massive MIMO Channel Transfer Function}
The time-variant channel transfer function is obtained as the Fourier transform of $h_{qp}(t,\tau)$ in \eqref{eq:CIR} wrt $\tau$ as
\begin{equation}
\begin{split}
H_{qp}(t,f) =&  \sum_{\ell=1}^{\mathcal{L}} c_\ell \lim_{N_\ell \to \infty} \frac{1}{\sqrt{N_\ell}} \\
&\times \sum_{n=1}^{N_\ell} a_{p,n\ell}  b_{q,n\ell} \: e^{j ( 2\pi f_{n,\ell} t + \theta_{n\ell} ) } e^{-j2\pi f \tau_{qp,n\ell}}.
\label{eq:H}
\end{split}
\end{equation}
Substituting (\ref{eq:delay}) into (\ref{eq:H}) and rearranging terms,
\begin{equation}
\begin{split}
H&_{qp}(t,f) =  \sum_{\ell=1}^{\mathcal{L}} c_\ell \lim_{N_\ell \to \infty} \frac{1}{\sqrt{N_\ell}} \\
&\times \sum_{n=1}^{N_\ell} {(a_{p,n\ell}  b_{q,n\ell})}^{\left(1+\frac{f}{f_0}\right)} \: e^{j ( 2\pi f_{n\ell} t + \theta_{n\ell} ) } e^{-j2\pi f \tau_{0,\ell}}
\label{eq:TF}
\end{split}
\end{equation}
where $f_0$ is the carrier frequency, i.e., $c_0 = \lambda_0 f_0$. Note that in conventional MIMO channel models, the term $\left(1+f/f_0\right)$ is usually approximated as 1 but, as it will be shown, the additional term $f/f_0$ introduced by the delay drift over the array has an impact in the correlation properties of the channel in the frequency domain. 
\section{Statistical Properties of the Channel Model}
\label{sec:stat_prop}
The STFCF of the CIR is defined as $\Gamma_{qp,q'p'}(\Delta t, f, \nu) = \text{E}[ H_{qp}^*(t,f) H_{q'p'}(t+\Delta t,f+\nu) ]$, with $\text{E}{[\cdot]}$ denoting the expectation operator. In this case, it can be calculated as
\begin{equation}
\begin{split}
\Gamma_{qp,q'p'}&(\Delta t,f, \nu) = \sum_{\ell=1}^{\mathcal{L}} c^2_\ell  e^{-j2\pi \nu \tau_{0,\ell}} \times \\
& \lim_{N_\ell \to \infty} \frac{1}{N_\ell}
\sum_{n=1}^{N_\ell} \text{E}\left[ a^2_{pp',n\ell} b^2_{qq',n\ell} e^{j 2\pi f_{n\ell}\Delta t}
 \right]_{\alpha^R_{n\ell}}
\label{eq:STFCF}
\end{split}
\end{equation}
where
\begin{eqnarray}
a_{pp',n\ell} = e^{j \frac{\pi}{c_0} \cos(\alpha_{n\ell}^T-\beta_T)  \left[ \Delta_{pp'} (f_0+f) + \delta_p \nu \right]} \label{eq:STFCF_a} \\
b_{qq',n\ell} = e^{j \frac{\pi}{c_0} \cos(\alpha_{n\ell}^R-\beta_R) \left[ \Delta_{qq'} (f_0+f) +  \delta_q\nu  \right]} \label{eq:STFCF_b}
\end{eqnarray}
with $\Delta_{pp'} =  (p-p') \delta_T$ and $\Delta_{qq'} = (q-q') \delta_R$ denoting the relative distance between $A^T_p$ and $A^T_{p'}$ and that between $A^R_q$ and $A^R_{q'}$, respectively. Note that it has been assumed uncorrelated scattering in the derivation of \eqref{eq:STFCF}, i.e., it has been considered that multipath components with different delays are uncorrelated. 
Also, notice in \eqref{eq:STFCF_a} and \eqref{eq:STFCF_b} the cross-product of the absolute and relative parameters, e.g., $\Delta_{qq'} (f_0+f)$ and $\delta_p  \nu$. The first product indicates that the space correlation function (SCF) depends on the frequency $f$ and the second that the frequency correlation function (FCF) depends on the antenna index, i.e., the absolute position over the array. Hence, the delay drift considered here results in cross non wide-sense stationary (WSS) properties of the channel along the array and in the frequency domain. These cross non-stationary properties will be studied in the following sections.
\subsection{Frequency-Variant SCF}
\label{sec:stat_prop_SCF}
The SCF of the $\ell$th path is calculated by setting the parameters $\nu=0$ and $\Delta t=0$ in \eqref{eq:STFCF}--\eqref{eq:STFCF_b} as 
\begin{equation}
\rho_{\ell}(\delta_T, \delta_R, f)=\sum_{\ell=1}^{\mathcal{L}} c^2_\ell \: \hat{r}_{\ell}(\delta_T, \delta_R,f) 
\label{eq:SCF_el}
\end{equation}
where the term $\hat{r}_{\ell}(\delta_T, \delta_R,f)$ denotes the SCF associated to the $\ell$th path and it is obtained as
\begin{equation}
\begin{split}
\hat{r}_{\ell}&(\delta_T, \delta_R,f) = \lim_{N_\ell \to \infty} \frac{1}{N_\ell} \times  \\
&\sum_{n=1}^{N_\ell} \text{E}\left[
 e^{j\frac{2\pi(f_0+f)}{c_0}  \left[ \Delta_{pp'}\cos(\alpha_{n\ell}^T-\beta_T)  +  \Delta_{qq'}\cos(\alpha_{n\ell}^R-\beta_R) \right] }
 \right]\!.
\label{eq:scf_l_sum}
\end{split}
\end{equation}
In the limit $N_\ell \to \infty$, $\hat{r}_{\ell}(\delta_T, \delta_R,f)$ approximates \cite{Patzold2012}
\begin{equation}
\begin{split}
\hat{r}_{\ell}(\delta_T, \delta_R,f)=& \int_{-\pi}^{\pi} e^{j\frac{2\pi}{c_0} (f_0+f) \Delta_{pp'} \cos(\alpha_{\ell}^T-\beta_T) }\\
& \times e^{j\frac{2\pi}{c_0} (f_0+f) \Delta_{qq'}\cos(\alpha_{\ell}^R-\beta_R) } p_{\alpha_\ell^R}(\alpha_\ell^R) d\alpha_\ell^R
\label{eq:scf_l_int}
\end{split}
\end{equation}
where the AOA $\alpha_{n,\ell}^R$ has been substituted by $\alpha_\ell^R$ with probability density function $p_{\alpha_\ell^R}(\alpha_\ell^R)$. Notice that the term $c_0/(f_0+f)$ in \eqref{eq:scf_l_sum} and \eqref{eq:scf_l_int} represents the wavelength at frequency $f_0+f$. Since the spatial correlation between different links decreases with the distance between antennas normalized to the wavelength, positive (negative) values of $f$ increase (decrease) correlation between links. Note that this frequency dependence applies to arrays irrespective of their dimensions. Nonetheless, this effect might be of particular interest for millimeter-Wave (mm-Wave) communication systems that are expected to operate using large bandwidths. 

When the AoA follows a von Mises distribution, closed-form expressions can be obtained for the one-side SCF. For instance, the receive-side SCF in \eqref{eq:scf_l_int} can be obtained using \cite[Eq. 3.338-4]{Grad} as
\begin{equation}
\begin{split}
\hat{r}_{\ell}(0, \delta_R,f) =& \frac{1}{I_0(\kappa)} I_0 \left( \left\{  
\kappa_\ell^2 - \left( \frac{2\pi f'\Delta_{qq'}}{c_0} \right)^2 \right. \right. \\
&\left. \left.+ j4\kappa \frac{\pi f' \Delta_{qq'}}{c_0}  \cos( \beta_R - m_{\alpha,\ell}^R)
\right\}^{1/2} \right) 
\label{eq:SCF_closed}
\end{split}
\end{equation}
where $f' = f_0+f$. Clearly, for narrowband systems where $f/f_0\ll 1$, then $f' \approx f_0$ in the previous expression, which reduces to the conventional receive-side SCF \cite{Patzold2012}, i.e., 

\begin{equation}
\begin{split}
\hat{r}_{\ell}(0, \delta_R,f) \approx& \frac{1}{I_0(\kappa)} I_0 \left( \left\{  
\kappa_\ell^2 - \left( \frac{2\pi\Delta_{qq'}}{\lambda_0} \right)^2 \right. \right. \\
&\left. \left.+ j4\kappa \frac{\pi \Delta_{qq'}}{\lambda_0}  \cos( \beta_R - m_{\alpha,\ell}^R)
\right\}^{1/2} \right).
\label{eq:SCF_closed_prev}
\end{split}
\end{equation}

\subsection{Array-Variant FCF}
\label{sec:stat_prop_FCF}
For large arrays and conventional bandwidths, the approximation $1+f/f_0\approx 1$ is still valid, but the terms $\delta_p\nu$ and $\delta_q\nu$ in \eqref{eq:STFCF_a} and \eqref{eq:STFCF_b} cannot be neglected. Thus, the FCF obtained as $R_{qp}(\nu) = \Gamma_{qp,qp}(0,0,\nu)$ depends on the position along the arrays as
\begin{equation}
\begin{split}
R_{qp} (\nu) =&
\sum_{\ell=1}^{\mathcal{L}} c_\ell^2  e^{-j2\pi \nu \tau_{0,\ell}} \times \\
&\lim_{N_\ell \to \infty} \frac{1}{N_\ell} 
\sum_{n=1}^{N_\ell} \text{E}\left[ a^2_{pp,n\ell} b^2_{qq,n\ell}
 \right].
\label{eq:FCF}
\end{split}
\end{equation}
Notice that the FCF is expressed as the product of two terms as $R_{qp} (\nu) = r (\nu)\cdot \tilde{r}_{qp}(\nu)$. The outer term in \eqref{eq:FCF} is the FCF of a tapped-delay Elliptical channel model in conventional MIMO systems, i.e., 
\begin{equation}
\begin{split}
r(\nu) =\sum_{\ell=1}^{\mathcal{L}} c_\ell^2  e^{-j2\pi \nu \tau_{0,\ell}}.
\label{eq:FCF_conv}
\end{split}
\end{equation}
On the other hand, the inner term in \eqref{eq:FCF} describes the frequency correlation introduced by the $\ell$th path. Thus, the path-level FCF is defined here as
\begin{equation}
\begin{split}
\tilde{r}_{qp,\ell}(\nu) &=
 \lim_{N_\ell \to \infty} \frac{1}{N_\ell} \times \\
&  \sum_{n=1}^{N_\ell} \text{E}\left[  e^{j\frac{2\pi \nu}{c_0} \left[\delta_p \cos(\alpha_{n\ell}^T-\beta_T) + \delta_q \cos(\alpha_{n\ell}^R-\beta_R) \right] }
\right].
\label{eq:FCF_nel}
\end{split}
\end{equation}
In the limit as $N_\ell \to \infty$, $\tilde{r}_{qp,\ell}(\nu,\delta_T, \delta_R)$ is given by \cite{Patzold2012}
\begin{equation}
\begin{split}
\tilde{r}_{qp,\ell}(\nu) = \hspace{-0.2cm}
 \int_{0}^{2\pi}  \hspace{-0.4cm} p_{\alpha_\ell^R}(\alpha_\ell^R)
 e^{j \frac{2\pi\nu}{c_0} \left[\delta_p \cos(\alpha_{\ell}^T\!-\beta_T) + \delta_q \cos(\alpha_{\ell}^R\!-\beta_R)\right]} \dif \alpha_\ell^R  .
\end{split}
\label{eq:FCF_el}
\end{equation}

Surprisingly, according to (\ref{eq:FCF_nel}) and (\ref{eq:FCF_el}), two signals of different frequencies bounced by the same ellipse of scatterers, i.e., rays from the same path, are not necessarily frequency-correlated. This implies that the FCF of a single-path model is not constant, unlike conventional MIMO models. This new effect can be explained as different rays bounced by the same ellipse but different scatterers travel different distances for antenna-elements of the array that are located far from the focus of the ellipse. We will show that this difference cannot be neglected when large-scale arrays are considered.

In some cases, it is possible to find the relationship between the length of the array and the decorrelation caused by the delay drift. For instance, let us consider the uplink of a massive MIMO system in which the base station (receiver) is composed of many antennas and the users (transmitters) are equipped with just a few antennas, i.e., $(M_T-1)\delta_T\nu/c_0 \ll 1$. 
Thus, it can be seen in (\ref{eq:FCF_el}) that the FCF of the $\ell$th path is constant as long as $\delta_q \cdot \nu = C_\ell$, with $C_\ell$ denoting a positive real constant. 
Consequently, the bandwidth where the path-level FCF is over a threshold (typically 0.5), i.e., the single-path coherence bandwidth $B_{\text{c},\ell}$, is inversely proportional to the antenna position over the array, i.e, $B_{\text{c},\ell}(q) = C_\ell/\delta_q$. Clearly, the constant $C_\ell$ depends on the correlation threshold considered and the specific characteristics of the scatterers such as their distribution in the angular domain.

When the AoA follows the von Mises distribution, it is possible to obtain the path-level FCF in \eqref{eq:FCF_el} using a similar procedure to the one used to obtain \eqref{eq:SCF_closed}, i.e., 
\begin{equation}
\begin{split}
\tilde{r}_{q,\ell}(\nu) =& \frac{1}{I_0(\kappa)} I_0 \left( \left\{  
\kappa^2 - \left( \frac{2\pi \delta_q \nu}{c_0}  \right)^2 \right. \right. \\
&\left. \left. \hspace{1.4cm}+ j4\kappa \frac{\pi \delta_q \nu}{c_0}  \cos( \beta_R - m_{\alpha,\ell}^R)
\right\}^\frac{1}{2} \right).
\label{eq:FCF_el_vm}
\end{split}
\end{equation}
Obvioulsy, for signals of the same frequency ($\nu=0$), the correlation is equal to 1 no matter the position over the array. 

\subsection{Array-Variant Average Power Delay Profile}
We will now study the array-variant average PDP for a multi-confocal Elliptical channel model and the first two central moments of the delay, i.e., the mean delay defined as $\tau_{\um,qp}=\text{E}{[\tau_{qp}]}$ and the delay spread defined as $\tau_{\text{rms},qp}=\sqrt{\text{E}{[\tau_{qp}^2]}-\tau_{\um,qp}^2}$. The PDP can be obtained as the inverse Fourier transform of the FCF as
\begin{equation}
\begin{split}
S_\tau(\tau) =& \int_{-\infty}^{\infty} 
                      R_{qp} (\nu) e^{j2\pi\nu\tau} \dif \nu.
\end{split}
\end{equation}
Note that $R_{qp} (\nu)$ in \eqref{eq:FCF} is the product of two functions. Then, applying the convolution theorem of the Fourier transform, it can be demonstrated that the PDP can be expressed as a linear combination of the path-level PDPs as  
\begin{equation}
\begin{split}
S_{\tau,qp}(\tau) = \sum_{\ell=1}^{\mathcal{L}} c_\ell^2 \: S_{\tau,qp,\ell}(\tau-\tau_{0,\ell})
\label{eq:}
\end{split}
\end{equation}
where $S_{\tau,qp,\ell}(\tau)$ is the inverse Fourier transform of $\tilde{r}_{qp,\ell}(\nu)$ in \eqref{eq:FCF_el}. For simplicity in the following analysis, we will assume here that only the Rx incorporates a large number of antennas. The path-level PDP can be obtained as 
\[
S_{\tau,q,\ell}(\tau) =
\begin{cases}
\frac{p_{\alpha_\ell^R}(\alpha_{\ell,1}^R) + p_{\alpha_\ell^R}(\alpha_{\ell,2}^R) }{\lvert \tau_q \rvert \sqrt{1 - \left(\tau/\tau_q\right)^2}}	&	  \lvert \tau \rvert < \tau_q \\
0 & \text{otherwise}\\
\end{cases}
\]
where $\alpha_{\ell,1}^R= \beta^R+\arccos(-\tau/\tau_q)$ and $\alpha_{\ell,2}^R= \beta^R-\arccos(-\tau/\tau_q)$. The mean delay is obtained as the weighted sum of the path-level mean delays as 
\begin{equation}
\begin{split}
\bar{\tau}_{q} =& \int_{-\infty}^{\infty} \!\!\!\!\!\! \tau S_{\tau,q}(\tau) \dif \tau \\
=& \sum_{\ell=1}^{\mathcal{L}} c_\ell^2 \: \int_{\tau_{0,\ell}-\tau_p}^{\tau_{0,\ell}+\tau_p}\!\!\!\!\! \tau S_{\tau,q,\ell}(\tau-\tau_{0,\ell}) \dif \tau \\
=& \sum_{\ell=1}^{\mathcal{L}} c_\ell^2 \: (\tau_{0,\ell} - \bar{\tau}_{\ell,q}).
\label{eq:mean_delay}
\end{split}
\end{equation}
where $\bar{\tau}_{\ell,q}$ denotes array-variant path-level mean delay drift. 
When the AOA follows the von Mises distribution with given mean direction of arrival and angular spread, the path-level mean delay $\bar{\tau}_{\ell,q}$ can be computed as
\begin{equation}
\begin{split}
\bar{\tau}_{\ell,q} = \int_{\tau_{0,\ell}-\tau_q}^{\tau_{0,\ell}+\tau_q}\!\!\!\!\! \tau S_{\tau,q,\ell}(\tau) \dif \tau  
= \frac{ \tau_q \cos( \beta^R-m_{\alpha,\ell}^R) I_1(\kappa_\ell)}{I_0(\kappa_\ell)}.
\label{eq:mean_delay_l}
\end{split}
\end{equation}
Although \eqref{eq:mean_delay_l} indicates that the path-level mean delay drifts over the array, the total mean delay $\bar{\tau}_{q}$ may remain independent of the antenna element depending on the angular distribution of the scatterers for every path. For instance, when most paths are concentrated in the angular domain in clusters, i.e., $\kappa_\ell \to \infty$, and they satisfy $\beta^R-m_{\alpha,\ell}^R= 0$ or $\beta^R-m_{\alpha,\ell}^R= \pi$ for $\ell=1 \dots \mathcal{L}$, the mean delay drift is maximized in absolute value and it tends to $|\tau_q|$. On the other hand, if we consider the mean AOA of these clusters distributed as $m_{\alpha,\ell}^R \sim \mathcal{U}(0,2\pi]$, the mean delay drift is reduced to zero as the lead-lag effect of signals coming from different directions balances positive and negative delays' drifting. 
For the von Mises distribution with given mean AOA and angular spread, the total delay spread can be obtained as 
\begin{equation}
\begin{split}
\tau_{\text{rms},q} =& \int_{-\infty}^{\infty} \!\!\!\!\!\! (\tau-\bar{\tau}_{q})^2 S_{\tau,q}(\tau) \dif \tau \\
=& \sum_{\ell=1}^{\mathcal{L}} c_\ell^2 \left\{ 
		\left(\tau_{0,\ell}-\bar{\tau}_{q}\right)^2	- 2\left(\tau_{0,\ell} - \bar{\tau}_{q}\right)\bar{\tau}_{\ell,q}  \right. \\ 
		&
		 - \frac{\tau_q^2}{\kappa_\ell I_0(\kappa_\ell)} \left[ \frac{\kappa_\ell \left[ I_0(\kappa_\ell) + I_2(\kappa_\ell) \right] \cos^2( \beta^R-m_{\alpha,\ell}^R)}{2} \right.  \\
			 & \left. \left. + I_1(\kappa_\ell) \sin^2( \beta^R-m_{\alpha,\ell}^R)   \right]  
		 \right\}^{1/2}.
\label{eq:rms_delay}
\end{split}
\end{equation}
Similarly, the distribution of $m_{\alpha,\ell}^R$ determines the drift of the delay spread over the array. Besides, it can be seen from \eqref{eq:mean_delay_l} and \eqref{eq:rms_delay} that a uniform distribution of the path-level AOA, i.e., $\kappa_\ell=0$, leads to zero-drift of the mean delay $\bar{\tau}_{q}$ and it maximizes the delay spread drift $\tau_{\text{rms},q}$. On the other hand, a very concentrated distribution, i.e., $\kappa_\ell \to \infty$, maximizes the mean delay drift for a given $m_{\alpha,\ell}^R$, but it tends to reduce the delay spread drift to zero.

\subsection{Limits of the Elliptical GBSM for Massive MIMO Systems}
One of the most important features of the elliptical GBSM lies upon describing {\it independent time dispersive and frequency dispersive} channels. In this kind of channels, the distributions of AOA and delays are independent, i.e., $\alpha_\ell^T = \alpha^T$, $\alpha_\ell^R = \alpha^R$, and $\tau_{n\ell}=\tau_\ell$. In conventional MIMO systems, this assumption is equivalent to consider that the STFCF can be separated into the product of the FCF and the space-time correlation function (STCF), which represents an important simplification of the channel. 
However, since the delay and the AOA are intrinsically related in arrays of large dimensions as indicated by (\ref{eq:delay}), such channels are necessarily dependent time dispersive and frequency dispersive. This can be seen by dropping the path index $\ell$ on the right-side of (\ref{eq:STFCF}), i.e.,
\begin{equation}
\begin{split}
\Gamma_{qp,q'p'}(\Delta t, f,\nu)=&\underbrace{\lim_{N \to \infty} \frac{1}{N} 
\sum_{n=1}^{N} \text{E}\left[ a^2_{pp',n} b^2_{qq',n} \: e^{j 2\pi f_{n}\Delta t}
 \right]}_\text{STCF and path-level FCF} \\ 
 & \times \underbrace{\sum_{\ell=1}^{\mathcal{L}} c^2_\ell   e^{-j2\pi \nu \tau_{0,\ell}}}_\text{FCF}.
\end{split}
\end{equation}
Since the terms $a^2_{pp',n}$ and $b^2_{qq',n}$ depend on the frequency separation $\nu$, the STFCF cannot be separated as intended. Consequently, the delay drift over large arrays breaks the assumption of time dispersive and frequency dispersive independence.
Notice that for narrowband systems, i.e., $f/f_0\ll 1$, and small antenna arrays, i.e., $(M_T-1)\delta_T\nu/c_0 \ll 1$, the STFCF is separable and the channel becomes independent time dispersive and frequency dispersive as expected.

For the von Mises distribution of the AOA, it can be readily seen in \eqref{eq:FCF_el_vm} that the iso-correlation curves satisfy the equation $\delta_q\cdot \nu = C$ with $C$ denoting a real positive constant. From \eqref{eq:FCF_el_vm}, the constant $C$ can be calculated as 
\begin{equation}
C = \frac{c_0}{2\pi}  \left(  \kappa^2 - \left[ I_0^{-1}( \rho I_0(\kappa) ) \right]^2  \right)^{1/2}
\end{equation}
with $I_0^{-1}(\cdot)$ the inverse of $I_0(\cdot)$ and $\rho$ the correlation threshold considered in the coherence bandwidth $B_{\text{c}}$.  For instance, for $\kappa=0$ and $\rho=0.5$, then $C = 1.936$ GHz. This value of $C$ can be interpreted as the maximum bandwidth for which the elliptical channel model could be used to simulate a two-antenna ULA MIMO system without introducing additional frequency decorrelation at the extremes of the ULA. Obviously, for a 100-antenna ULA, this bandwidth decreases by a factor 100, i.e., it becomes 19.36 MHz.

\begin{figure}[t]
\centering
 \includegraphics[width=0.45\textwidth]{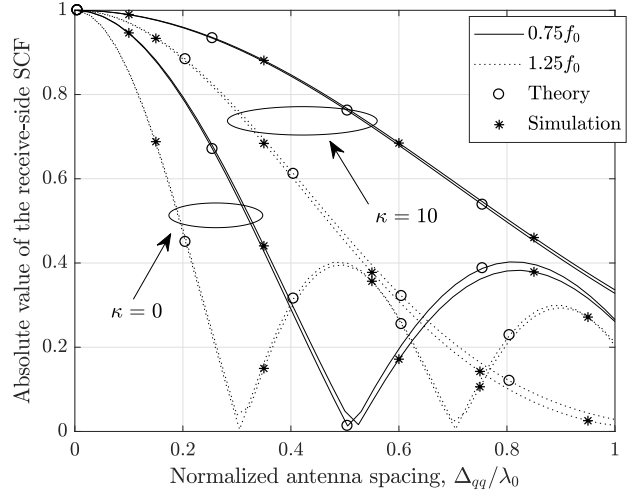} 
  \caption{Absolute value of the SCF for a single ellipse channel model for different values of $\kappa $ ($f_0=2$ GHz, $M_T=1$, $M_R=2$, $\delta_R=\lambda_0/2$, $\beta_R-m^R_\alpha = \pi/2$).}
\label{fig:SCF_k_cuts}
\end{figure}

\section{Results and Analysis}
\label{sec:results}
In this section, we will study the statistical properties obtained in Section~\ref{sec:stat_prop} for both narrowband (single-) and wideband (multi-)confocal Elliptical models. Due to the high dimensionality of \eqref{eq:scf_l_int} and \eqref{eq:FCF_el}, the study of the SCF and FCF only in the Rx's side will be shown here by setting $\delta_T=0$, i.e., considering that the Tx is composed by a single antenna. In the following, theoretical results refer to the numerical evaluation of the corresponding expressions in Section~\ref{sec:stat_prop} and the Monte Carlo method is used to obtain simulation results.

\subsection{Narrowband Model Results (Single-Ellipse)}
To illustrate the dependence of the SCF $\hat{r}_{\ell}(0, \delta_R,f)$ in \eqref{eq:SCF_closed} on the frequency $f$, Fig.~\ref{fig:SCF_k_cuts} shows two samples of its absolute value at 1.5 GHz and 2.5 GHz. The SCF is asymmetrical around $f=0$, as noted in Section~\ref{sec:stat_prop_SCF}, because the spatial correlation increases at lower frequencies due to the reduction of the normalized distance between antennas $\delta_R/\lambda_0$ and vice versa. Moreover, for low values of the frequency, i.e., $|f|\ll f_0$, the SCF can be considered frequency-independent. Although the effect is relatively small, it can be observed that the decorrelation effect is more pronounced for higher values of $\kappa$, i.e., for lower angular spreads.

\begin{figure}[b]
\centering
 \includegraphics[width=0.45\textwidth]{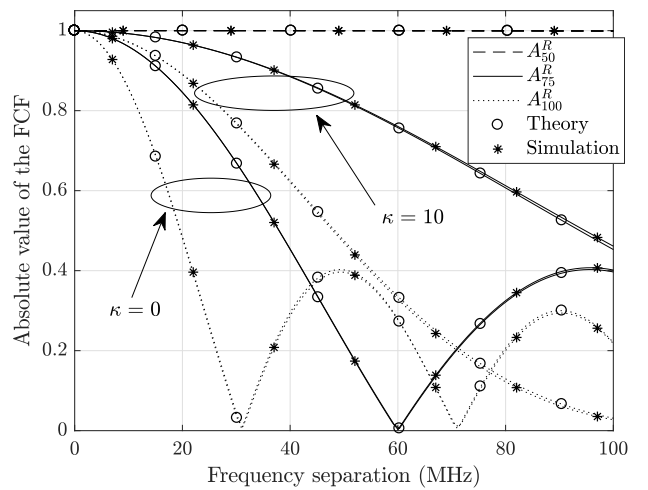} 
  \caption{Absolute value of the FCF along the array for a single-path channel for different values of $\kappa $ ($f_0=2$ GHz, $\delta_R=\lambda_0/2$, $M_T=1$, $M_R=100$, $\beta_R-m^R_\alpha = \pi/2$).}
\label{fig:FCF_k_cuts}
\end{figure}

In Fig.~\ref{fig:FCF_k_cuts}, the array-variant FCFs of a single-path $\tilde{r}_{q,\ell}(\nu)$ in \eqref{eq:FCF_el_vm} for different values of the von Mises $\kappa$ parameter are presented. As the ULA is about 50$\lambda_0$ long, the extremes of the array are approximately located at -25$\lambda_0$ and 25$\lambda_0$. The FCFs are obtained at the receiving antennas $A^R_{50}$, $A^R_{75}$, and $A^R_{100}$, i.e, at the center, midpoint, and one extreme of the antenna-array. 
As it would be expected for a narrowband channel model, the FCF is almost constant over the whole band for small-size arrays, i.e., around $A^R_{50}$. However, signals of different frequencies become less correlated as the antenna-element considered is further away from the array's center or ellipse's focus. This clearly indicates that a single-ellipse channel model becomes increasingly frequency-selective over the array. Unlike wideband channel models, this phenomenon is caused by the delay drift over the array and not by multiple paths (ellipses) in the channel. Note also that the frequency decorrelation over the array is more pronounced for low values of $\kappa$, i.e., for larger values of the angular spread. Thus, these results also indicate a dependency of the FCF with respect to the path-level distribution of the AOA.

\subsection{Wideband Model Results (multi-confocal Elliptical)}
In this section we will consider a wideband channel model by employing multiple ellipses whose associated delays are randomly distributed according to an exponential distribution, i.e., $f_\tau(x) = 1/\tau_\text{rms}\exp[-x/\tau_\text{rms}]$ for $x \in [0, \infty)$. As the impact of the delay drift is more important in environments whose delay spreads are similar to the maximum delay drift over the array, hence we have used $\tau_\text{rms}=30$ ns, which is slightly higher than the typical values for indoor environments, e.g., propagation scenarios A1 (indoor office) and B3 (indoor hotspot), and for some outdoor environments, e.g., propagation scenario D1 (rural macro-cell), in WINNER-II models \cite{WINNERII-models}. The simulation results were obtained by generating $10^3$ exponentially distributed delays and $10^2$ scatterers per delay with AOAs following a von Mises distributions of random parameters. The PDP is obtained through the empirical probability distribution function of \eqref{eq:delay} and the FCF as its Fourier transform. The gains associated to every path (ellipse) are deterministic variables meeting the condition $\sum_{\ell=1}^{\mathcal{L}} c_\ell ^2 = 1$, e.g., $c_\ell=1/\sqrt{\mathcal{L}}$.

\begin{figure}[t]
\centering
 \includegraphics[width=0.45\textwidth]{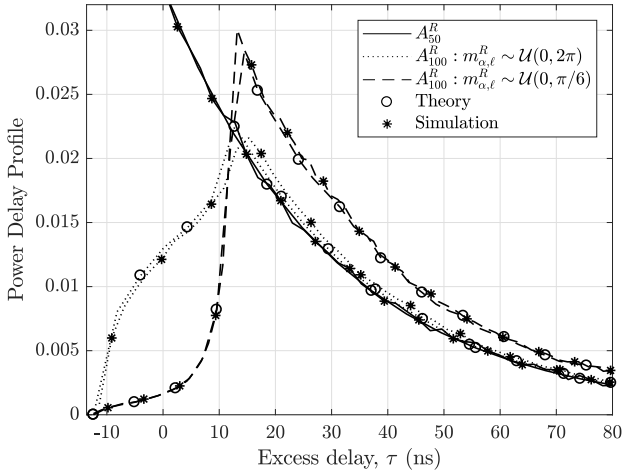} 
  \caption{Array-variant PDP for a wideband multi-confocal Elliptical channel model ($f_0=2$ GHz, $M_T=1$, $M_R=100$, $\delta_R=\lambda_0/2$, $\beta_R=0$, $\tau_\text{rms} = 30$ ns).}
\label{fig:PDP_multi_ellipse}
\end{figure}

In Fig.~\ref{fig:PDP_multi_ellipse}, the PDPs of the channel at the center ($A^R_{50}$) and at one extreme of the array ($A^R_{100}$) are presented for different spreads of the clusters' mean AOAs. In the two cases studied, the mean AOAs are distributed as $m_{\alpha,\ell}^R\sim \mathcal{U}(0,2\pi)$ and $m_{\alpha,\ell}^R\sim \mathcal{U}(0,\pi/6)$. Moreover, the concentration parameters are distributed as $\kappa_\ell \sim \mathcal{U}(0,10)$, corresponding to a range of angular spreads from 18 to 104 degrees approximately.

First, it can be observed that the PDPs at $A^R_{100}$ are shifted and spread and, as a result, they cover negative values of the excess delay. As delays are measured relative to the center of the array, negative values indicate that there are scattered signals reaching one extreme of the array before reaching its center. For high spreads of $m_{\alpha,\ell}^R$, whereas the PDPs at $A^R_{100}$ and $A^R_{50}$ are significantly different in the region $\tau<|\tau_q|$ approximately, both are very similar for higher values of $\tau$. 
The similarities for $\tau>|\tau_q|$ are due to the balancing effect of negative and positive drifts of the signals scattered by different ellipses, i.e., signals of low delay drifted towards positive delays are balanced by those of higher delay drifted towards negative values. This does not occur for the lowest part of the PDP ($\tau<|\tau_q|$) due to a boundary effect, i.e., lowest-delay signals drifted toward negative delays cannot be balanced by those of lower delay as they do not exist. The similarity between the PDPs for $\tau>|\tau_q|$ is only possible because it has been assumed that $m_{\alpha,\ell}^R\sim \mathcal{U}(0,2\pi)$. Under this assumption, the mean delay of the channel is not modified as it was shown in Section~\ref{sec:stat_prop}. 
On the other hand, when $m_{\alpha,\ell}^R$ are distributed over a smaller interval, there is a net drift of the PDP over the array and the spreading is less pronounced, as it can be seen in Fig.~\ref{fig:PDP_multi_ellipse}. This is caused by the concentration of clusters towards specific directions, which eliminates the balance effect of positive and negative drifts over the array.

\begin{figure}[t]
\centering
 \includegraphics[width=0.45\textwidth]{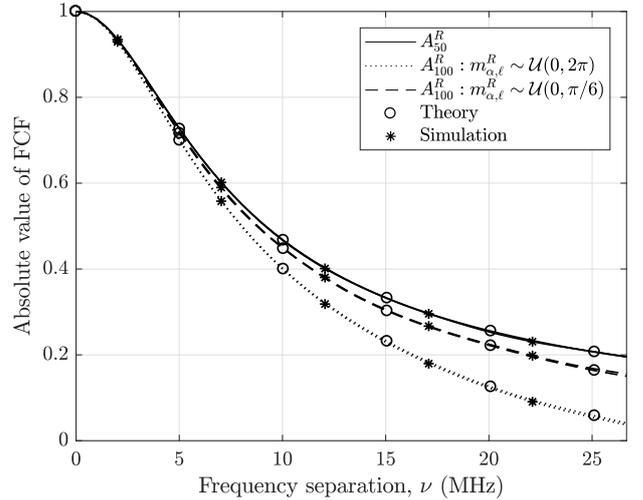} 
  \caption{Array-variant FCF for a wideband multi-confocal Elliptical channel model ($f_0=2$ GHz, $M_T=1$, $M_R=100$, $\delta_R=\lambda_0/2$, $\beta_R=0$, $\tau_\text{rms} = 30$ ns).}
\label{fig:FCF_multi_ellipse}
\end{figure}

In Fig.~\ref{fig:FCF_multi_ellipse}, the FCFs of the channel at the center and at one extreme of the array ($A^R_1$) are presented for different spreads of the clusters' mean AOA $m_{\alpha,\ell}^R$. 
As it can be seen, the differences between the FCF at the center and at one extreme of the array are accentuated when $m_{\alpha,\ell}^R\sim \mathcal{U}(0,2\pi)$ and they are smaller when $m_{\alpha,\ell}^R\sim \mathcal{U}(0,\pi/6)$. In agreement with the analysis of \eqref{eq:mean_delay}--\eqref{eq:rms_delay} and Fig.~\ref{fig:PDP_multi_ellipse}, a larger spread of $m_{\alpha,\ell}$ results in a larger spread of the PDP over the array, hence the reduction in the coherence bandwidth of the channel. On the other hand, as it has been argued, a small spread of $m_{\alpha,\ell}$ leads to a drift of the PDP, which is reflected in the phase of the FCF, but it cannot be noticed in its absolute value. 

\section{Conclusion}
\label{sec:conclusions}
In this paper, the impact of the delay drift in the statistical properties of both narrowband and wideband massive MIMO channel models has been studied. It has been shown that delay drifts can cause non-stationary properties of the channel in the frequency and spatial domains. It has also been demonstrated that the correlation between signals of different frequencies depends on the position along the array and that is largely affected by the size of the array in massive MIMO systems. Closed-form expressions of the FCF and SCF for a specific massive MIMO GBSM have been obtained and analyzed. Moreover, we have shown that the spread of the clusters' mean AOA determines whether the PDP is subject to a net delay drift or to a spreading effect.
Finally, we can conclude that extending conventional MIMO GBSMs to model massive MIMO channels requires to consider the delay drift as a source of decorrelation. 

\section*{Acknowledgment}
The authors gratefully acknowledge the support of this work from the EU H2020 5G Wireless project (Grant No. 641985), the EU FP7 QUICK project (Grant No. PIRSES-GA-2013-612652), the EPSRC TOUCAN project (Grant No. EP/L020009/1), and the EU H2020 RISE TESTBED project (Grant No. 734325).

\bibliographystyle{IEEEtran}
\bibliography{IEEEabrv,Bibliography} 

\begin{thebibliography}{10}
\providecommand{\url}[1]{#1}
\csname url@samestyle\endcsname
\providecommand{\newblock}{\relax}
\providecommand{\bibinfo}[2]{#2}
\providecommand{\BIBentrySTDinterwordspacing}{\spaceskip=0pt\relax}
\providecommand{\BIBentryALTinterwordstretchfactor}{4}
\providecommand{\BIBentryALTinterwordspacing}{\spaceskip=\fontdimen2\font plus
\BIBentryALTinterwordstretchfactor\fontdimen3\font minus
  \fontdimen4\font\relax}
\providecommand{\BIBforeignlanguage}[2]{{%
\expandafter\ifx\csname l@#1\endcsname\relax
\typeout{** WARNING: IEEEtran.bst: No hyphenation pattern has been}%
\typeout{** loaded for the language `#1'. Using the pattern for}%
\typeout{** the default language instead.}%
\else
\language=\csname l@#1\endcsname
\fi
#2}}
\providecommand{\BIBdecl}{\relax}
\BIBdecl

\bibitem{Persson2012}
D.~Persson, B.~K. Lau, and E.~G. Larsson, ``{Scaling up MIMO},'' \emph{IEEE
  Signal Process. Mag.}, vol.~30, no.~1, pp. 40--60, Jan. 2013.

\bibitem{Larsson2013}
E.~G. Larsson, O.~Edfors, F.~Tufvesson, and T.~L. Marzetta, ``{Massive MIMO for
  next generation wireless systems},'' \emph{IEEE Commun. Mag.}, vol.~52,
  no.~2, pp. 186--195, Feb. 2013.

\bibitem{Wang14ComMag}
C.~X. Wang, F.~Haider, X.~Gao, X.~H. You, Y.~Yang, D.~Yuan, H.~M. Aggoune,
  H.~Haas, S.~Fletcher, and E.~Hepsaydir, ``Cellular architecture and key
  technologies for 5g wireless communication networks,'' \emph{IEEE Commun.
  Mag.}, vol.~52, no.~2, pp. 122--130, Feb. 2014.

\bibitem{Björnson2016}
E.~Bj{\"o}rnson, E.~G. Larsson, and T.~L. Marzetta, ``{Massive MIMO: ten myths
  and one critical question},'' \emph{IEEE Commun. Mag.}, vol.~54, no.~2, pp.
  114--123, Feb. 2016.

\bibitem{Tse2005}
D.~Tse and P.~Viswanath, \emph{{Fundamentals of wireless communication}},
  1st~ed.\hskip 1em plus 0.5em minus 0.4em\relax Cambridge: University Press,
  2005.

\bibitem{Gao2012}
X.~Gao, F.~Tufvesson, O.~Edfors, and F.~Rusek, ``{Measured propagation
  characteristics for very-large MIMO at 2.6 GHz},'' in \emph{Proc. IEEE
  ASILOMAR'12}, Pacific Grove, USA, Nov. 2012, pp. 295--299.

\bibitem{Gao2013}
X.~Gao, F.~Tufvesson, and O.~Edfors, ``{Massive MIMO channels - measurements
  and models},'' in \emph{Proc. IEEE ASILOMAR'13}, Pacific Grove, USA, Nov.
  2013, pp. 280--284.

\bibitem{Li2015}
W.~Li, L.~Liu, C.~Tao, Y.~Lu, J.~Xiao, and P.~Liu, ``{Channel measurements and
  angle estimation for massive MIMO systems in a stadium},'' in \emph{Proc.
  IEEE ICACT'15}, Seoul, South Korea, July 2015, pp. 105--108.

\bibitem{Payami2012}
S.~Payami and F.~Tufvesson, ``{Channel measurements and analysis for very large
  array systems at 2.6 GHz},'' in \emph{Proc. IEEE EUCAP'12}, Prague, Czech
  Republic, Mar. 2012, pp. 433--437.

\bibitem{Gao2015}
X.~Gao, O.~Edfors, F.~Rusek, and F.~Tufvesson, ``{Massive MIMO performance
  evaluation based on measured propagation data},'' \emph{IEEE Trans. Wireless
  Commun.}, vol.~14, no.~7, pp. 3899--3911, July 2015.

\bibitem{Gao2015a}
X.~Gao, O.~Edfors, F.~Tufvesson, and E.~G. Larsson, ``{Massive MIMO in real
  propagation environments: do all antennas contribute equally?}'' \emph{IEEE
  Trans. Commun.}, vol.~63, no.~11, pp. 3917--3928, Nov. 2015.

\bibitem{SCM}
``{Spatial channel model for multiple-input multiple-output (MIMO)
  simulations},'' 3GPP T.S. 25.996, Tech. Rep. v11.0.0, 2012.

\bibitem{WINNERII-models}
K.~Kyosti \emph{et~al.}, ``{WINNER II channel models: part I channel models},''
  IST-4-027756 WINNER II, Tech. Rep. D1.1.2 V1.2, 2007.

\bibitem{Verdone2011}
R.~Verdone and A.~Zannella, \emph{{Pervasive mobile and ambient wireless
  communications -- the COST action 2100}}, 1st~ed.\hskip 1em plus 0.5em minus
  0.4em\relax Springer, 2011.

\bibitem{IMT-A}
``{Guidelines for evaluation of radio interface technologies for
  IMT-Advanced},'' ITU-R M.2135, Switzerland, Tech. Rep., 2008.

\bibitem{Wu2014}
S.~Wu, C.~X. Wang, E.~H.~M. Aggoune, M.~M. Alwakeel, and Y.~He, ``{A
  non-stationary 3-D wideband twin-cluster model for 5G massive MIMO
  channels},'' \emph{IEEE J. Sel. Areas Commun.}, vol.~32, no.~6, pp.
  1207--1218, June 2014.

\bibitem{Lopez16}
C.~F. L{\'o}pez, C.~X. Wang, and R.~Feng, ``A novel 2d non-stationary wideband
  massive mimo channel model,'' in \emph{IEEE CAMAD'16}, Oct 2016, pp.
  207--212.

\bibitem{Lopez2018}
C.~L{\'o}pez and C.-X. Wang, ``Novel 3d non-stationary wideband models for
  massive mimo channels,'' \emph{IEEE Trans. Wireless Commun. accepted for
  publication}.

\bibitem{Wu2015a}
S.~Wu, C.-X. Wang, H.~Haas, H.~Aggoune, M.~M. Alwakeel, and B.~Ai, ``{A
  non-stationary wideband channel model for massive MIMO communication
  systems},'' \emph{IEEE Trans. Wireless Commun.}, vol.~14, no.~3, pp.
  1434--1446, Mar. 2015.

\bibitem{Quadriga2014}
S.~Jaeckel, L.~Raschkowski, K.~Börner, and L.~Thiele, ``{QuaDRiGa: A 3-D
  multi-cell channel model with time evolution for enabling virtual field
  trials},'' \emph{IEEE Trans. Antennas Propag.}, vol.~62, no.~6, pp.
  3242--3256, June 2014.

\bibitem{Metis15}
L.~Raschkowki, ``{METIS channel models},'' Tech. Rep., 2015.

\bibitem{Wang2016b}
C.-X. Wang, S.~Wu, L.~Bai, X.~You, J.~Wang, and C.-L. I, ``{Recent advances and
  future challenges for massive MIMO channel measurements and models},''
  \emph{Sci. China Inf. Sci.}, vol.~59, no.~2, pp. 1--16, Feb. 2016.

\bibitem{Patzold2012}
M.~P{\"a}tzold, \emph{{Mobile radio channels}}, 2nd~ed.\hskip 1em plus 0.5em
  minus 0.4em\relax West Sussex: John Wiley \& Sons, 2012.

\bibitem{Grad}
I.~S. Gradshteyn and I.~M. Ryzhik, \emph{{Table of Integrals, Series, and
  Products}}, 8th~ed.\hskip 1em plus 0.5em minus 0.4em\relax London, UK:
  Academic Press, 2014.

\end{thebibliography}

\end{document}